\documentclass[twocolumn,showpacs,preprintnumbers,amsmath,amssymb,pre,floatfix,showkeys]{revtex4}

\usepackage[english]{babel}
\usepackage[latin2]{inputenc}
\usepackage[T1]{fontenc}

\usepackage{ae,aecompl}
\usepackage{enumerate}
\usepackage{epsfig}
\usepackage{graphics}
\usepackage{graphicx}
\usepackage{amsmath}
\usepackage{amssymb}
\usepackage{pifont}
\usepackage[sort&compress]{natbib}
\usepackage{wasysym}

\def\ccc#1;#2{\left\langle #1 \left\vert #2 \right.\right\rangle}
\def\ev #1{\left\langle #1 \right\rangle}

\begin{document}

\preprint{}
\title{The dynamics of traded value revisited}
\author{Zolt\'an Eisler}
\email{eisler@maxwell.phy.bme.hu}
\author{J\'anos Kert\'esz}
\altaffiliation[Also at ]{Laboratory of Computational Engineering, Helsinki
University of Technology, Espoo, Finland} \affiliation{Department of Theoretical
Physics, Budapest University of Technology and Economics, Budapest, Hungary}
\date{\today}


\begin{abstract}
We conclude from an analysis of high resolution NYSE data that the distribution of the traded value $f_i$ (or volume) has a finite variance $\sigma_i$ for the very large majority of stocks $i$, and the distribution itself is non-universal across stocks. The Hurst exponent of the same time series displays a crossover from  weakly to strongly correlated behavior around the time scale of $1$ day. The persistence in the strongly correlated regime increases with the average trading activity $\ev{f_i}$ as $H_i=H_0+\gamma\log\ev{f_i}$, which is another sign of non-universal behavior. The existence of such liquidity dependent correlations is consistent with 
the empirical observation that $\sigma_i\propto\ev{f_i}^\alpha$, where $\alpha$ is a non-trivial, time scale dependent exponent.
\end{abstract}

\maketitle

The recent years have seen a number of important contributions of physics to various areas, among them to finance \cite{bouchaud.book, stanley.book}. The application of physical concepts often seems well suited to the analysis of financial time series, however, it is not without caveats. Often, the theoretical background of these methods is deeply rooted in physical laws that -- naturally -- do not apply to stock markets. In particular, observations regarding power laws \cite{futurepl}, universality \cite{culturecrash}, and other empirical regularities \cite{gallegati.etal} are often criticized. We carried out a thorough study of the traded value per unit time \cite{eisler.non-universality, eisler.sizematters, eisler.unified} and have arrived at the result that some earlier conclusions have to be modified. Here we present an analysis of some new data, which supports our earlier findings.

The paper is organized as follows. Section \ref{sec:intro} introduces notations. Section \ref{sec:value} shows that the distribution of traded volume/value is not universal, and it is not in the Levy stable regime as suggested by Ref. \cite{gopi.volume}. Section \ref{sec:correl} shows, that traded value displays only weak correlations for time scales shorter than one day. On longer horizons there is stronger persistence whose degree depends logarithmically on the liquidity of the stock. Finally, Section \ref{sec:alpha} surveys the concept of fluctuation scaling, shows how it complements the observed liquidity dependence of correlations, and how those two form a consistent scaling theory.

\section{Notations and data}
\label{sec:intro}
For a fixed time window size $\Delta t$, let us denote the total traded value
of the $i$th stock at time $t$ by 
\begin{equation}
f_i^{\Delta t}(t) = \sum_{n, t_i(n)\in [t, t+\Delta t]} V_i(n),
\label{eq:flow}
\end{equation} 
where $t_i(n)$ is the time when the $n$th transaction of the $i$th stock
takes place. Tick-by-tick data are denoted by
$V_i(n)$, this is the value traded in transaction $n$, calculated as the
product of the price and the traded volume.

Since price changes very little from trade to trade while variations of trading volume are much faster, 
the fluctuations of the traded value $f_i(t)$ are basically determined by those of traded volume. Price merely acts as a weighting factor that enables one to compare different stocks, while this also automatically corrects the data for stock splits and dividends. The correlation properties and the normalized distribution are nearly indistinguishable between traded volume and traded value.

This study is based on the complete Trades and Quotes database of New York
Stock Exchange for the period $1994-1995$.

Note that throughout the paper we use $10$-base logarithms.

\section{Traded value distributions revisited}
\label{sec:value}

In this section, we first revisit the analysis done in Ref. \cite{gopi.volume}. That work finds that the cumulative distribution function of traded volume for time windows of $\Delta t = 15$ minutes decays as a power-law with a tail exponent $\lambda = 1.7 \pm 0.1$ for a wide range of stocks. This is the so called \emph{inverse half cube law}, and it can be written as
\begin{equation}
{\mathbb P}_{\Delta t}(f) \propto f^{-(\lambda + 1)},
\label{eq:pl}
\end{equation}
where $\mathbb P_{\Delta t}$ is the probability density function of the same quantity.

\begin{figure*}[tb]
\centerline{\includegraphics[height=205pt,trim=0 0 10 0]{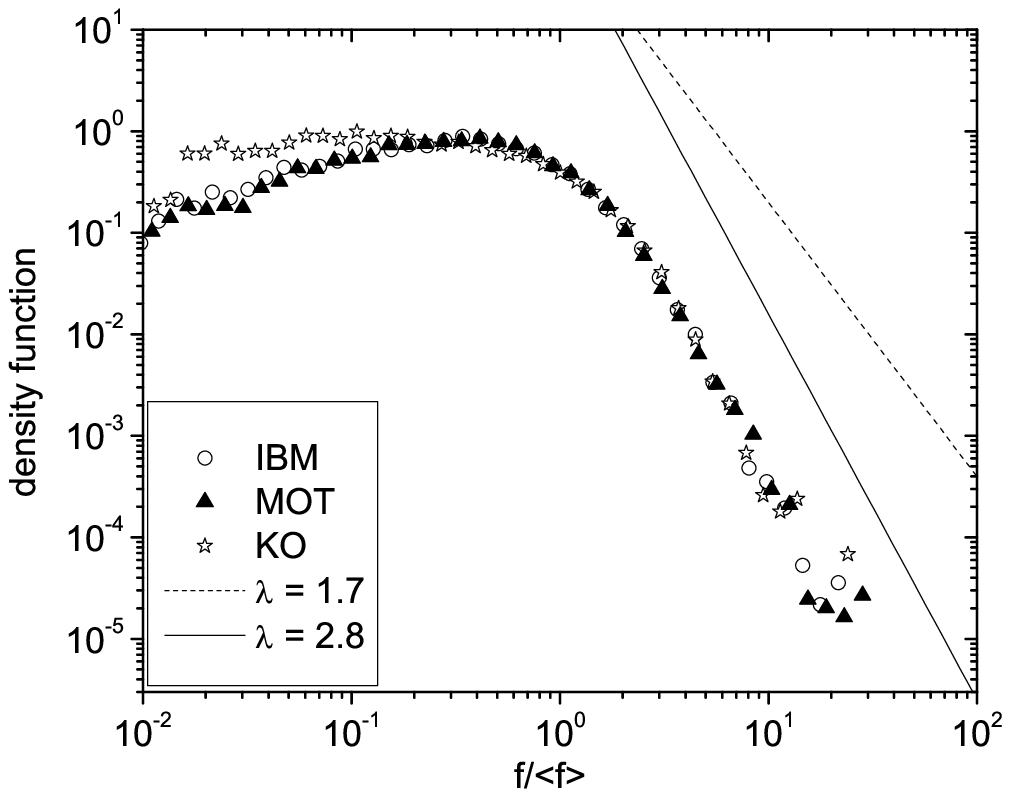}\includegraphics[height=205pt]{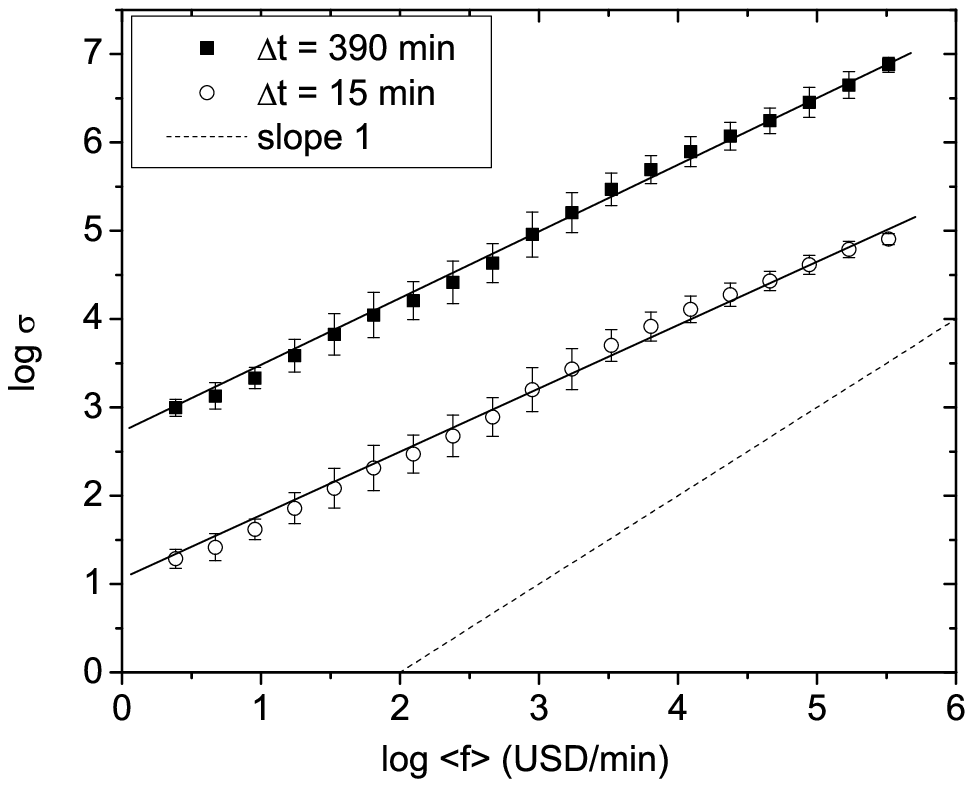}}
\caption{(left) The distribution of traded value in $\Delta t = 15$ min time
windows, normalized by the mean. The plot displays three example stocks
for the period $1994-1995$. The dashed and solid diagonal lines represent
power-law distributions with tail exponents $\lambda = 1.7$ and $2.8$, respectively. (right) The standard deviation $\sigma_i$ of trading activity plotted versus the mean $\ev{f_i}$ of the same quantity. Both for $\Delta t = 15$ min, and  $\Delta t = 390$ min $\approx 1$ trading day, there is a power law relationship with slopes $\alpha(15$ min$)=0.74$, and $\alpha(390$ min$)=0.78$. A linear proportionality would support the existence of a universal distribution. However, the sublinear scaling relationship suggests that relative fluctuations in trading activity are smaller for stocks with higher average liquidity. \emph{Note}: For better visibility, stocks were binned according to $\ev{f}$ and their $\log \sigma$ was averaged. The error bars correspond to the characteristic range within the bins.}
\label{fig:distrib}
\end{figure*}

The estimation of tail exponents is often difficult due to poor statistics of rare events, large stock-to-stock variations and the presence of correlations. For the same $1994-1995$ period of data and the same $15$ minute time window certain stocks have $\lambda$ values significantly higher than $1.7$ [see Fig. \ref{fig:distrib}(left)]. The tails of these distributions can be fitted by a power law over an order of magnitude, for the top $3-10\%$ of the events. The exponent $\lambda$ is around $2.8$ for these examples. The question arises: Which value (if any) is correct?

In order to address this question we carried out a systematic investigation comprising the $1000$ stocks with the highest total traded value in the TAQ database. We used variants of Hill's method \cite{hill,alves} to estimate the typical tail exponent, see Ref. \cite{eisler.sizematters} for details. The results of this Section are summarized in Table \ref{tab:DETRlambda94-95}. Note that in all cases the $U$-shaped intraday pattern of trading activity was removed.

Most descendants of Hill's method, including the ones applied here, contain a free parameter, namely the fraction $p$ of top events to be considered to belong to the tail of the distribution (see Ref. \cite{alves} and refs. therein). According to Fig. \ref{fig:distrib}(left) this should be set around $p\approx 3-10\%$.

First, let us follow the methodology of Ref. \cite{gopi.volume}. In that paper, the authors first they deduct the mean from the time series by taking $f_i(t)-\ev{f_i}$, where $\ev{\cdot}$ denotes time averaging. Then this series is used to estimate value of $\lambda$ by applying Hill's method \cite{gopi.personal}. The choice $p=0.03$ provides results in line with Ref. \cite{gopi.volume}, for $\Delta t = 15$ min time windows one finds $\lambda = 1.67 \pm 0.20$. There are several issues with this approach:
\begin{enumerate}
\item $p$ is a parameter that can be chosen arbitrarily. With the variation of $p$ the same procedure can produce estimates ranging from $\lambda = 1.1\pm 0.2$ ($p=0.10$) to $\lambda = 2.15 \pm 0.2$ ($p=0.005$). 
\item The transformation significantly decreases the estimates of $\lambda$, down to the range of Levy stable distributions ($\lambda < 2$). Estimates for the untransformed data are given in Table \ref{tab:DETRlambda94-95} for comparison.
\end{enumerate}

It is simple to show, that the first issue emerges, i.e. the estimates systematically depend on $p$, when one applies Hill's method to a finite sample from a distribution of the form
\begin{equation}
{\mathbb P}_{\Delta t}(f) \propto (f+f_0)^{-(\lambda + 1)},
\label{eq:pl2}
\end{equation}
where $f_0$ is a non-zero constant. The transformation to $f_i(t)-\ev{f_i}$ does not resolve the problem, but biases the estimates further.

Instead, to correct for these biases one can (i) either find the proper constant $f_0$, remove it from the data, and apply Hill's estimator afterwards (ii) or apply the estimator of Fraga-Alves \cite{alves}, which is insensitive to such shifts. Both of these estimates were found to be significantly higher \footnote{The Fraga-Alves estimator converges very slowly, and it underestimates the actual values of $\lambda$ from small samples. Its estimates can be interpreted as lower bounds for $\lambda$.}: $\lambda > 2$, see Table \ref{tab:DETRlambda94-95}. The methods are described in detail in Ref. \cite{eisler.sizematters}.

The two corrected estimators show a strong tendency of increasing $\lambda$ with
increasing $\Delta t$. Monte Carlo simulations on surrogate datasets
show that this is beyond what could be explained by decreasing sample
size. For distributions with $\lambda < 2$ increasing window size should result in a convergence to the corresponding Levy distribution, and the measured $\lambda$'s should be
independent of $\Delta t$. Only when $\lambda > 2$ can the measured effective
value of $\lambda$ systematically increase with $\Delta t$. For the $95\%$ of the stocks the increasing tendency is observed and for a window size $\Delta t = 1$ day the respective $\lambda$'s are greater than $2$. These are strong indications that the distributions are not in the Levy stable regime, and thus the second moment exists.

Note that our calculations \emph{assume} that the variable is asymptotically distributed as \eqref{eq:pl2} and do not \emph{prove} it. Still, the existence of the second moment is guaranteed by the absence of convergence to a Levy distribution. Consequently, it is possible to define the Hurst exponent for $f_i(t)$.

\begin{table*}[tbp]
	\centering
		\begin{tabular}{c||c||c||c|c||c|}
		$\Delta t$ & Hill's $\lambda$ ($p=0.06$) & Ref. \cite{gopi.volume}, $p=0.03$ & Shifted Hill's $\lambda$ & $f_0/\ev{f}$ & Fraga Alves  ($p=0.1$)\\
		\hline\hline
		$1$ min & $1.43 \pm 0.09$ & $1.45\pm0.10$ & $2.15\pm0.15$ & $3.0$ & $1.98 \pm 0.25$ \\
		$5$ min & $1.56 \pm 0.13$ & $1.55\pm0.15$ & $2.29 \pm 0.25$ & $2.8$ & $2.04 \pm 0.25$ \\	
		$15$ min & $1.71 \pm 0.20$ & $1.67\pm0.20$ & $2.55 \pm 0.35$ & $2.8$ & $2.1 \pm 0.3$ \\	
		$60$ min & $2.06 \pm 0.30$ & $1.90\pm0.25$ & $2.85 \pm 0.45$ & $1.8$ & $2.1 \pm 0.4$ \\			
		$120$ min & $2.3 \pm 0.4$ & $2.0\pm0.3$ & $3.15 \pm 0.70$ & $1.6$ & $2.1 \pm 0.4$ \\	
		$390$ min & $2.7 \pm 0.6$ & $2.1\pm0.5$ & $3.7 \pm 0.9$ & $1.2$ & no estimate \\	
		\hline
		\end{tabular}
	\caption{Median of the tail exponents of traded value calculated by four methods for $1994-1995$. The width of the distributions is given with the half distance of the $25\%$ and $75\%$ quantiles.}
	\label{tab:DETRlambda94-95}
\end{table*}

Regardless of the absence of the convergence to Levy stability there are qualitative similarities in the shape of the traded value distributions of various stocks [cf. Fig. \ref{fig:distrib}(left)]. Nevertheless, the existence of a universal distribution can be rejected by a simple test \footnote{Similar techniques were used in Refs. \cite{lillo.variety, eisler.sizematters} to show non-universality in the distribution of returns and intertrade times.}.

If the form of the normalized distribution was universal, then the ratio of the standard deviation and the mean would have to obey $\sigma_i/\ev{f_i}=h$, where $h$ is a constant independent of the stock. Equivalently, a relationship
\begin{equation}
	\sigma_i \propto \ev{f_i}^\alpha
\label{eq:alpha_first}
\end{equation}
would have to hold with an exponent $\alpha = 1$, at least on average. Even though one finds a monotonic dependence between the two quantities [as shown in Fig. \ref{fig:distrib}(right)], the exponent is significantly less than $1$. This means, that the ratio $\sigma/\ev{f}$ decreases with growing $\ev{f}$, i.e., the normalized distribution of $f$ is narrower for larger stocks, so their trading exhibits smaller relative fluctuations. We will return to this observation in Section \ref{sec:alpha}.

\section{Non-universality of correlations in traded value time series}
\label{sec:correl}

One of the classical tools of both financial analysis and physics is the measurement of the correlation properties of time series \cite{bouchaud.book, stanley.book, tumminello}. In particular, scaling methods \cite{dfa} have a long tradition in the study of physical systems, where the Hurst exponent $H_i$ is often calculated. For the traded value time series $f_i^{\Delta t}(t)$ of stock $i$ this is defined as
\begin{equation}
\label{eq:hurst}
\sigma_i^2(\Delta t) = \ev{\left [f_i^{\Delta t}(t)-\ev{f_i^{\Delta t}(t)} \right ]^2}\propto\Delta t^{2H_i},
\end{equation}
Note that it follows from the results of Section \ref{sec:value} that the variance on the left hand side exists regardless of stock and for any window size $\Delta t$.

\begin{figure}[tb]
\centerline{\includegraphics[height=200pt]{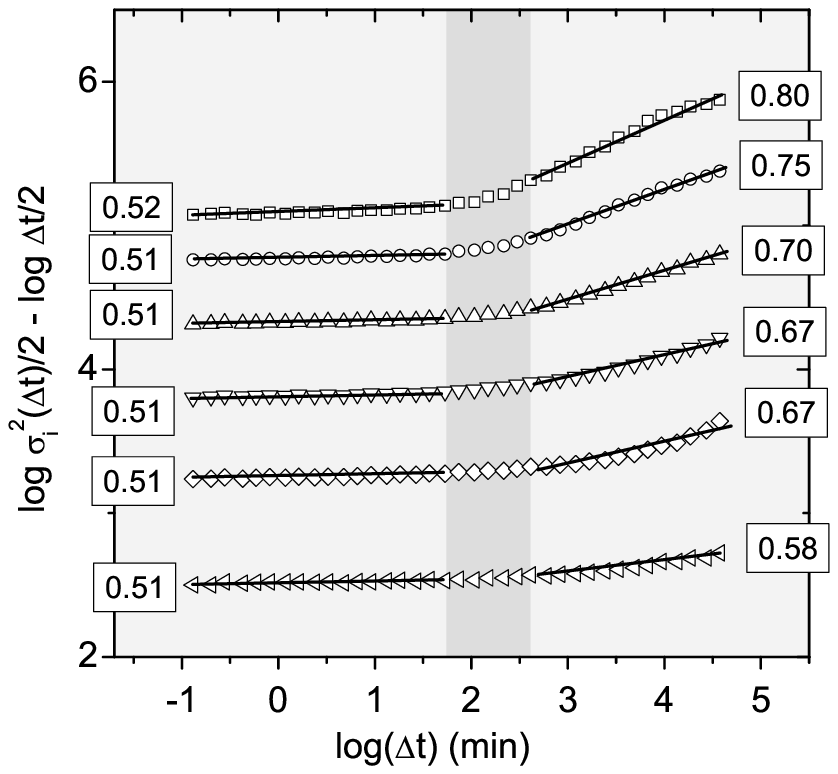}}
\caption{The normalized variance $\frac{1}{2}\log \sigma_i^2(\Delta t)-\frac{1}{2}\log \Delta t$ for the six groups of companies, with average traded values $\ev{f}\in[0,10^4)$, $\ev{f}\in[10^4,10^5)$, \dots, $\ev{f}\in[10^8,\dots)$ USD/min, increasing from bottom to top. A horizontal line would mean the absence of autocorrelations in the data. Instead, one observes a crossover phenomenon in the regime $\Delta t = 60-390$ mins, indicated by darker background. Below the crossover all stocks show very weakly correlated behavior, $H^-\approx 0.5$. Above the crossover, the strength of correlations, and thus the slope corresponding to $H^+-\frac{1}{2}$, increases with the liquidity of the stock. The asymptotic values of $H^{\pm}$ are indicated in the plot.}
\label{fig:scavg}
\end{figure}

The measurements were carried out for all $2474$ stocks that were continuously
available on the market during $1994-1995$ \footnote{For a similar analysis of the years $2000-2002$, see Ref. \cite{eisler.sizematters}.}. Then we sorted the stocks into $6$ groups according to the order of magnitude of their average traded value: $0\leq \ev{f}\leq 10^4$, $10^4\leq \ev{f}\leq 10^5$, \dots, $10^8\leq\ev{f}$, all values in USD/min. Finally we averaged $\sigma_i^2(\Delta t)$ within each group. The obtained scaling plots are shown in Fig. \ref{fig:scavg}.

\begin{figure}[tb]
\centerline{\includegraphics[height=205pt]{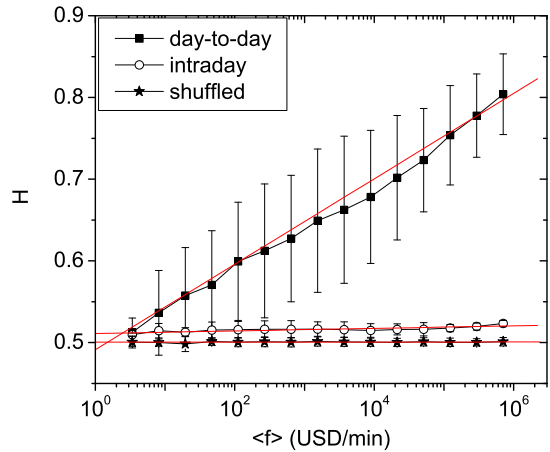}}
\caption{Value of the Hurst exponents of traded value for the time period $1994-1995$. For short time windows (O, $\Delta t < 60$ min) all signals are nearly uncorrelated, $H^-\approx 0.51 - 0.53$. The fitted slope is $\gamma^-=0.00\pm 0.01$. For larger time windows ($\blacksquare$, $\Delta t > 390$ min) the strength of correlations depends logarithmically on the mean trading activity of the stock, 
$\gamma^+=0.053\pm 0.01$ for $1994-1995$. Shuffled data ($\bigstar$) display no correlations, thus $H_{\mathrm{shuff}} = 0.5$, which also implies $\gamma_\mathrm{shuff} = 0$. {\it Note}: Groups of stocks were binned, and their Hurst exponents were averaged. The error bars correspond to the standard deviations in the bins.}
\label{fig:hurst}
\end{figure}

All stocks display a crossover around window sizes of $\Delta t = 60-390$ min, and there are two sets of Hurst exponents: $H^-_i$ valid below, and $H^+_i$ above the crossover. These characterize the strength of intraday and long time correlations, respectively. The behavior on these two time scales is very different. 
\begin{enumerate}
	\item For intraday fluctuations, regardless of stock $H^-\approx 0.51-0.52$. This means that intraday fluctuations of traded value are nearly uncorrelated. 
	\item For long time fluctuations the data are correlated, but the strength of correlations depends strongly on the liquidity of the stock. As one moves to groups of larger $\ev{f}$, the strength of correlations ($H^+$) grows, up to $H^+\approx 0.8$.
	\item If one shuffles the time series, correlations are destroyed, and $H_{\rm shuff} = 0.5$.
\end{enumerate}

The same phenomenon can be characterized by directly plotting the dependence of $H^\pm$ on $\ev{f}$, as done in Fig. \ref{fig:hurst}. Such a dependence is well described by a logarithmic law:
\begin{eqnarray}
H_i^{\pm} = H_0^\pm + \gamma^\pm \log \ev{f_i},
\label{eq:hurst_scaling}
\end{eqnarray}
where $\gamma^-=0.00\pm0.01$, and $\gamma^+=0.053\pm0.01$. For the shuffled time series $\gamma_{\rm shuff}=0$.

These results indicate, at least in the case of traded value, the absence of universal behavior. Liquidity (or, analogously, company size) is a relevant quantity, which acts as a \emph{continuous} parameter of empirical observables, in particular the strength of correlations and the distribution of $f$. Related results can be found in Refs. \cite{eisler.liquidity, bonanno.dynsec, ivanov.itt, eisler.sizematters}.

\begin{figure}[tb]
\centerline{\includegraphics[width=255pt]{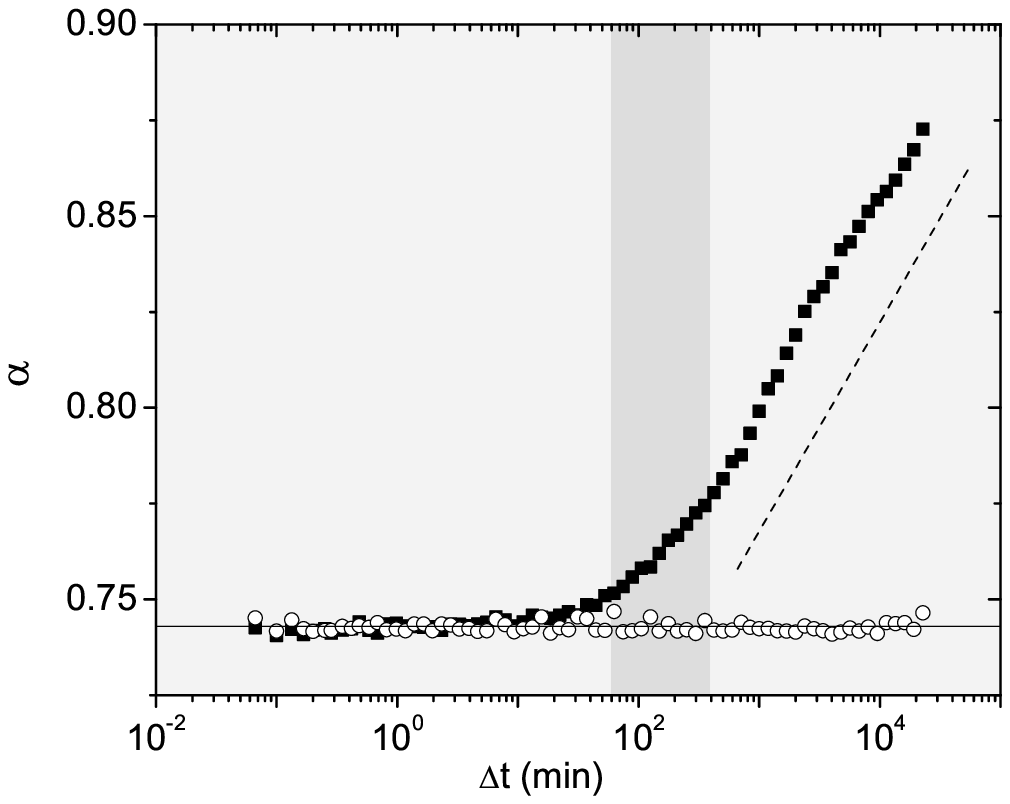}}
\caption{The dependence of the scaling exponent $\alpha$ on the window
size $\Delta t$ for the years $1994-1995$. The lighter shaded intervals have well-defined Hurst exponents
and values of $\gamma$, the crossover is indicated with a darker background. Without shuffling ($\blacksquare$)
there are two linear regimes: For shorter windows $\alpha = 0.74 \pm 0.02$, the slope is $\gamma^-=\gamma(\Delta t<60$
min$)=0.00\pm 0.01$ (solid line), while for longer windows $\alpha$ grows logaritmically, with a slope $\gamma^+=\gamma(\Delta t>390$ min$)=0.052\pm 0.01$ (dashed line). For shuffled data (O) the exponent is independent of window size, $\alpha (\Delta t)=0.74\pm0.02$.}
\label{fig:alpha}
\end{figure}

\section{Fluctuation scaling}
\label{sec:alpha}
Fluctuation scaling is a general phenomenon, observed in a wide range of complex systems \cite{barabasi.fluct, eisler.non-universality, eisler.unified}: A scaling law connects the standard deviation $\sigma_i$ and the average $\ev{f_i}$ of the same quantity. In the case of the trading activity of stocks we have already presented this result in Section \ref{sec:value} [cf. Eq. \eqref{eq:alpha_first}]. Now we give a more detailed discussion. 

Let us start from our observation that
\begin{equation}
\sigma_i(\Delta t) \propto \ev{f_i}^{\alpha (\Delta t)},
\label{eq:alpha}
\end{equation}
where the scaling variable is $\ev{f_i}$, or more appropriately the stock $i$, and $\Delta t$ is kept constant [see Fig. \ref{fig:distrib}(right)]. Notice that $\sigma_i(\Delta t)$ is the same as in the definition of the Hurst exponent in Eq. \eqref{eq:hurst}, where $i$ was constant and $\Delta t$ was varied.

In Eq. \eqref{eq:alpha} the window size $\Delta t$ is a free parameter. This scaling law persists for any $\Delta t$, but $\alpha$ strongly depends on its value, as shown in Fig. \ref{fig:alpha}. For small time windows (up to $60$ min), $\alpha(\Delta t)\approx 0.74$, then, after a crossover regime, when $\Delta t > 390$ min, there is a logarithmic trend. This can be summarized as
\begin{equation}
\alpha(\Delta t) = \alpha_0^\pm + \gamma^\pm\log \Delta t,
\label{eq:alpha_scaling}
\end{equation}
where $\cdot^\pm$ refers to the regimes $\Delta t < 60$ min and $\Delta t > 390$ min. The constants are $\alpha_0^- = 0.74$, $\gamma^- =0$, and $\gamma^+ = 0.052\pm 0.01$. For shuffled time series, $\alpha(\Delta t) = 0.74$ regardless of $\Delta t$, i.e., $\gamma_{\rm shuff}=0$.

A visual comparison of Figs. \ref{fig:scavg} and \ref{fig:alpha} reveals that the crossover in the behavior of $\alpha(\Delta t)$ and $H$ falls into the same interval. Moreover, when $\Delta t < 60$ min, both $\alpha(\Delta t)$ and $H^-(\ev{f})$ are constant. For $\Delta t > 390$ min, both $\alpha(\Delta t)$ and $H^+(\ev{f})$ vary logarithmically with their arguments (see Figs. \ref{fig:hurst} and \ref{fig:alpha}).

In order to better understand the connection between temporal correlations and fluctuation scaling, let us repeat here Eqs. \eqref{eq:hurst_scaling} and \eqref{eq:alpha_scaling}:
\begin{eqnarray}
	\alpha(\Delta t)=\alpha_0^\pm + \gamma^\pm \log \Delta t, \nonumber \\
	H_i=H_0^\pm + \gamma^\pm \log \ev{f_i}. \nonumber
\end{eqnarray}
Beyond the obvious symmetry of these two logarithmic laws, notice that the prefactors are equal: in both equations $\gamma^- \approx 0$ and $\gamma^+\approx 0.05$. 

It is easy to show \cite{eisler.unified} that none of this is a simple coincidence. If both fluctuation scaling and long range autocorrelations are present in data, there are only two possible ways for their coexistence: 
\begin{enumerate}
\item Correlations are homogeneous throughout the system, $H_i=H_0$, $\gamma = 0$, and $\alpha$ is independent of $\Delta t$. This is realized for $\Delta < 60$ min. For shuffled time series correlations are absent, thus such data also fall into this category.
\item Both the $H(\ev{f_i})$ and $\alpha(\Delta t)$ are logarithmic functions of their arguments with the same coefficient $\gamma^+$. This is realized for $\Delta > 390$ min.
\end{enumerate}

In other words the coexistence of the two scaling laws is so restrictive, that if the strength of correlations depends on $\ev{f}$ at all, then the realized logarithmic dependence is the only possible scenario.

\section{Conclusions}

In this paper, we analyzed the empirical properties of trading activity on the New York Stock Exchange. We showed that, in contrast to earlier findings, the distribution of traded value is not in the Levy stable regime, and is not universal. Traded value is nearly uncorrelated on an intraday time scale, while on daily or longer scales fluctuations show strong persistence, whose strength grows logarithmically with the liquidity of the stock. This effect is in harmony with findings on fluctuation scaling, a general scaling framework for complex systems. 

All our results imply, that the notion of universality must be used with extreme care in the context of financial markets, where the concepts and the theoretical background are radically different from those in physics. The liquidity of a stock strongly affects the distribution and the correlation structure of its trading activity. This dependence is continuous, which means the absence of universality classes in trading dynamics. The dynamical process responsible for such a dependence is yet to be identified.

Finally, we would like to make two remarks. Firstly, in Refs. \cite{gopi.volume} and \cite{gabaix.volatility} it is stated that the so called inverse half cubic law is observable not only on the $15$ minute level but also in the tick-by-tick data. Our analysis dealt with data aggregated for $1$ minute and more, and we showed that the assumption of a power law decay is not consistent with the inverse half cubic law for these cases.  Secondly, in Ref. \cite{gabaix.volatility} a footnote mentions the possibility of an exponential cutoff in the distribution. This assumption would influence the estimators strongly and we did not consider this case. We thank the anonymous referee for calling our attention to these points.

The authors thank Gy\"orgy Andor and \'Ad\'am Zawadowski for their help with the data.
ZE is grateful for the hospitality of l'Ecole de Physique des Houches.
JK is member of the Center for Applied Mathematics and Computational Physics, BME. 
Support by OTKA T049238 is acknowledged.

\bibliographystyle{apsrev}
\bibliography{Eisler}

\end{document}